\documentclass[a4paper, 11pt] {article}

\usepackage{amsmath}
\usepackage{amsfonts}
\usepackage{amsthm}
\usepackage[body={5.6in,8.75in}] {geometry}
\usepackage{cite}
\usepackage{graphicx}
\usepackage{indentfirst}
\usepackage{palatino}
\usepackage[onehalfspacing]{setspace}

\def \ba {\begin{eqnarray}}
\def \ea {\end{eqnarray}}
\def \bas {\begin{eqnarray*}}
\def \eas {\end{eqnarray*}}
\def \be {\begin{equation}}
\def \ee {\end{equation}}
\def \bes {\begin{equation*}}
\def \ees {\end{equation*}}
\def \bpm {\begin{pmatrix}}
\def \epm {\end{pmatrix}}
\def \bpm {\begin{pmatrix}}
\def \epm {\end{pmatrix}}
\def \bse {\begin{subequations}}
\def \ese {\end{subequations}}

\def \bar {\overline}
\def \bb {\mathbb}

\def \cal {\mathcal}
\def \hs {\hspace{5pt}}
\def \it {\textit}
\def \ni {\noindent}
\def \no {\nonumber}

\def \pa {\prime}
\def \pb {\prime \prime}

\def \rm {\mathrm}

\def \wh {\widehat}

\def \half {{\textstyle \frac{1}{2}}}

\begin{document}
\begin{titlepage}
\title{Phase space spinor amplitudes for spin-$\half$ systems}     
     
\author{P. Watson\footnote{{\em Email:} pw.cmp@optusnet.com.au} \hs and A.J. Bracken\footnote{{\em Email:} a.bracken@uq.edu.au}\\Centre for Mathematical Physics\\Department of Mathematics\\    
University of Queensland\\Brisbane 4072\\Queensland\\Australia}     
\date{} %\date{\normalsize\today}
\maketitle     

\begin{abstract}
\ni The concept of phase space amplitudes for systems with continuous degrees of freedom is generalized to finite-dimensional spin systems.
Complex amplitudes are obtained on both a sphere and a finite lattice, in each case enabling a more fundamental description of pure spin states than that previously given by Wigner functions.
In each case the Wigner function can be expressed as the star product of the amplitude and its conjugate, so providing a generalized Born interpretation of amplitudes that emphasizes their more fundamental status.
The ordinary product of the amplitude and its conjugate produces a (generalized) spin Husimi function.
The case of spin-$\half$ is treated in detail, and it is shown that phase space amplitudes on the sphere transform correctly as spinors under under rotations, despite their expression in terms of spherical harmonics.
Spin amplitudes on a lattice are also found to transform as spinors.
Applications are given to the phase space description of state superposition, and to the evolution in phase space of the state of a spin-$\half$ magnetic dipole in a time-dependent magnetic field.
\end{abstract}

\end{titlepage}

\setcounter{page}{2}

\section{Introduction}
Weyl quantization \cite{Weyl1950} is a one to one map between functions on a phase space $\Gamma$ and operators acting on a Hilbert space $\cal{H}$.
Over the years the Weyl-Wigner (WW) transform $\cal{W}$, which is the inverse of Weyl's quantization map, has been used to develop the phase space formulation of quantum mechanics that has successfully been applied to systems with continuous degrees of freedom \cite{Hillery1984, Lee1994}.

There have been several attempts to generalize the Weyl correspondence to include finite spin degrees of freedom within the WW formalism.
The treatments of deGroot and Suttorp \cite{deGroot1972} and O'Connell and Wigner \cite{Wigner1984} combined the continuous phase space picture with the finite spin degrees of freedom into a single WW transform. 
Buot \cite{Buot1974} formulated a discrete WW transform on a periodic lattice phase space; Chumakov \it{et al} \cite{Chumakov2000} defined a WW type quasi-probability function as a linear combination of spherical harmonics on the phase space of the sphere $\cal{S}^2$; and Berezin \cite{Berezin1974, Berezin1975} introduced a general method of quantization by representing functions on $\cal{S}^2$ in terms of covariant Q-symbols and contravariant P-symbols.

The lattice phase space approach was further generalized to an array of $N$ orthogonal states to obtain a discrete Wigner function.
For example Hannay and Berry \cite{Hannay1980} obtained a discrete Wigner function that resembled a delta-function on each of the $(2N)^2$ points of a $2N \times 2N$ lattice phase space indexed by the set of integers modulo $2N$.
Wootters \cite{Wootters1987} generated a general class of discrete Wigner functions (applicable to spin systems) by constructing an $N \times N$ lattice phase space indexed by the set of integers modulo $N$, where $N$ is a prime number, or a Cartesian product of lattices in the case of composite $N$.
Gibbons \it{et al} \cite{Gibbons2004} found a class of discrete Wigner functions on an $N \times N$ lattice phase space of a finite field composed of $N^k$ elements where $N$ is prime and $k \in \bb{Z}^+$. 
Other approaches that dealt with spin systems but from a different perspective are those of Chandler \it{et al} \cite{Chandler1992} who, motivated by the statistical characteristic function, generated three dimensional quasi-probability density functions for spin-$\half$ systems; and Leonhardt \cite{Leonhardt1996}, who applied precession tomography to spin systems to derive an expression for a discrete Wigner function.

Phase space amplitudes have been defined for continuous degrees of freedom \cite{TorresVega1993, Harriman1994, Wlodarz1994, Oliveira2004, deGosson2005, Smith2006,
Bracken2010}.
They provide representations of quantum state vectors rather than density operators, and as such are more fundamental objects than Wigner functions \cite{Bracken2010}.
In this paper we extend the concept of phase space amplitudes to finite-dimensional spin systems.
We begin with a brief review of phase space amplitudes for continuous degrees of freedom and then modify the main results to define phase space amplitudes for a spin $j$ system.
Expressions for spin amplitudes and their corresponding Wigner functions are obtained on both the sphere $\cal{S}^2$ and a lattice of dimension $(2j+1)^2$.
Our primary focus will be on spin-$\half$ amplitudes and explicit expressions for these will be given in terms of spherical harmonics and functions defined on each of $(2j+1)^2 = 4$ lattice points.
For convenience we consider all variables to be dimensionless and set Planck's constant $\hbar$ to unity.

\section{Phase space amplitudes}
In the phase space formulation of quantum mechanics, state vectors for a system with continuous degrees of freedom $(p,q)$ only are represented by complex phase space amplitudes \cite{TorresVega1993, Harriman1994, Wlodarz1994, Oliveira2004, deGosson2005, Smith2006,
Bracken2010}.
These are constructed from a pure state vector $|\psi\rangle$ by introducing a fixed ``window'' state $|\varphi\rangle$ of unit length, to form the outer (dyadic) product 
\be \label{dyad1}
\wh{\Psi} \equiv |\psi\rangle\langle\varphi|.
\ee
For each choice of $|\varphi\rangle$, a set of distinct amplitudes for variable $|\psi\rangle$ are then found as the images $\cal{W}(\wh{\Psi})$ under the WW transform,
\be
\cal{W}(\wh{\Psi})(p,q)/\sqrt{2\pi} = \Psi(p,q) =  \rm{Tr}\left(\wh{\Psi}\wh{\Delta}(p,q)\right)/\sqrt{2\pi}.
\ee
Here $\wh{\Delta}(p,q)$ is the Weyl-Wigner-Stratonovich (WWS) operator kernel \cite{Stratonovich1957}.

A phase space amplitude contains all essential information about a quantum state, and this representation of state vectors in phase space has many important properties.
For instance, phase space amplitudes can be used to calculate expectation values according to
\be \label{average1}
\langle\wh{A}\rangle = \rm{Tr}(\wh{A}\wh{\Psi}\wh{\Psi}^\dagger) \longrightarrow \int \bar{\Psi(p,q)} \; [(A \star \Psi)(p,q)] \; \rm{d}\Gamma, \qquad (\rm{d}\Gamma = \rm{d}p\rm{d}q),
\ee
and transition amplitudes
\be
\langle\psi_1|\psi_2\rangle = \int \bar{\Psi_1(p,q)} \Psi_2(p,q) \; \rm{d}\Gamma,
\ee
where in particular
\be
\langle\psi|\psi\rangle = \int \bar{\Psi(p,q)} \Psi(p,q) \; \rm{d}\Gamma = 1,
\ee
so that the quantity $|\Psi|^2$ can be regarded as a quasi-probability distribution over $\Gamma$.
It is remarkable \cite{Bracken2010} that $|\Psi|^2$ is the familiar Husimi distribution \cite{Husimi1940} if $|\varphi\rangle$ is chosen to be the ground state of the harmonic oscillator, and for other choices of the window state $|\varphi\rangle$, it is a generalized Husimi function \cite{Bracken2010}.
In (\ref{average1}), the associative and non-commutative star-product appears, and in terms of phase space functions $A(p,q)$ and $B(p,q)$, is given by  \cite{Moyal1949}
\ba \label{starint1}
(A \star B)(p,q) & = & \left(\frac{1}{\pi}\right)^2\int \rm{Tr}\left(\wh{\Delta}(p,q)\wh{\Delta}(p^{\pa},q^{\pa})\wh{\Delta}(p^{\pb},q^{\pb})\right) A(p^{\pa},q^{\pa}) B(p^{\pb},q^{\pb}) \no\\
& & \qquad\qquad\qquad\times\;\rm{d}p^{\pa}\rm{d}q^{\pa}\rm{d}p^{\pb}\rm{d}q^{\pb}.
\ea
The Wigner function $W(p,q)$ is defined for a pure state $|\psi\rangle$ as the image $\cal{W}(|\psi\rangle\langle\psi|)$ and is commonly expressed as \cite{Wigner1932}
\be
\cal{W}(|\psi\rangle\langle\psi|)(p,q)/2\pi \longleftrightarrow W(p,q) = \frac{1}{2\pi} \int e^{ipx} \bar{\psi(q + x/2)} \psi(q - x/2) \; \rm{d}x.
\ee

In analogy to the way that configuration space wave functions determine a probability density, phase space amplitudes generalize the Born interpretation by determining the Wigner function with the help of the star product
\be \label{born1}
W(p,q) = (\Psi \star \bar{\Psi})(p,q),
\ee
which emphasizes the more fundamental status of these amplitudes.

\section{Spin amplitudes on the sphere $\cal{S}^2$}
For each spin observable $\wh{J}_k, \; k = 1,2,3$, represented by $(2j+1) \times (2j+1)$ hermitian matrices with the fixed spin quantum number $j \in 0, 1/2, 1 \cdots$, the inverse of Weyl's quantization map assigns a real valued function on the phase space of the sphere $\cal{S}^2$.
However, pure spin states described in the Hilbert space $\cal{H}_j = \bb{C}^{2j+1}$ by eigenvectors $|j,m\rangle$ of $\wh{J}_3$ and $\wh{J}^2$,
\be
\wh{J}_3 |j,m\rangle = m |j,m\rangle, \; \wh{J}^2 |j,m\rangle = j(j+1) |j,m\rangle, \qquad m = j, j-1, \cdots -j,
\ee
do not have images defined on $\cal{S}^2$.

Within the framework of phase space amplitudes this problem is circumvented by first observing that spin states are vectors $|\psi\rangle$ with $2j + 1$ complex components $\psi_m = \langle j,m|\psi\rangle$ in $\cal{H}_j$ and second, in a manner similar to the continuous case, by introducing a fixed spin state $|\varphi\rangle$ which allows one to extend the amplitude operator of (\ref{dyad1}) to 
\be \label{ampop1}
\wh{\Psi} \equiv |\psi\rangle\langle\varphi| \longleftrightarrow (\psi_m\bar{\varphi}_{m^{\pa}}) = 
   \bpm
      \psi_j\bar{\varphi}_j     & \psi_j\bar{\varphi}_{j-1} & \cdots \\
      \psi_{j-1}\bar{\varphi}_j & \ddots                    & \\
      \vdots                    &                           & \psi_{-j}\bar{\varphi}_{-j}
   \epm
   .
\ee
Normalization of $|\varphi\rangle$ requires that
\be
\sum^j_{m=-j} |\varphi_m|^2 = 1.
\ee

A basic feature of Weyl quantization is the existence of an operator valued function or WWS kernel, the determination of which is essential for the definition not only of the Wigner functions on $\cal{S}^2$, as in the literature \cite{Chumakov2000}, but also of the spin amplitudes in which we are interested here.
A WWS kernel based on the axiomatic postulates of Stratonovich \cite{Stratonovich1957} and applicable to spin systems is known \cite{Varilly1989, Amiet1991, Heiss2000} and given by
\bse \label{kern1}
\ba
\wh{\Delta}^j(\theta,\phi) & = & \sum^j_{m^{\pa},m^{\pb}=-j} Z^j_{m^{\pa} m^{\pb}} |j,m^{\pb}\rangle\langle j,m^{\pa}| \\
Z^j_{m^{\pa} m^{\pb}} & = & \sqrt{\frac{4\pi}{2j+1}} \sum^{2j}_{l=0} \sum^l_{m=-l} \epsilon^j_l (-1)^{j-m^{\pa}} C^{\;j\quad\;\;l\;\;\;\;\;j}_{m^{\pb} \; -m^{\pa} \; m} \; Y_{l,m}(\theta,\phi),
\ea
\ese
where $C^{\;j\quad\;\;l\;\;\;\;\;j}_{m^{\pb} \; -m^{\pa} \; m}$ are the standard Clebsch-Gordan coefficients, $Y_{l,m}$ are the spherical harmonics and $\epsilon^j_l$ are constants such that $\epsilon^j_0 = 1 \;\rm{and}\; \epsilon^j_l = \pm 1$. 

It is now a straightforward matter using (\ref{ampop1}) and (\ref{kern1}), after making a suitable choice for the arbitrary fixed spin vector $|\varphi\rangle$, to define spin amplitudes on $\cal{S}^2$ as
\be \label{spamp1}
\Psi(\theta,\phi) = \rm{Tr}\left(\wh{\Psi}\wh{\Delta}^j(\theta,\phi)\right),
\ee
for each state $|\psi\rangle$ in (\ref{ampop1}).
Whatever choice of fixed spin vector $|\varphi\rangle$ is made in (\ref{ampop1}), the corresponding Wigner function on $\cal{S}^2$, which is defined as 
\be
W(\theta,\phi) = \rm{Tr}(|\psi\rangle\langle\psi| \wh{\Delta}^j(\theta,\phi)),
\ee
is then expressed in terms of the spin amplitude by the star product
\be \label{spamp1a}
W(\theta,\phi) = (\Psi \star \bar{\Psi)}(\theta,\phi).
\ee
There are various methods available to evaluate star products for spin systems \cite{Klimov2002, Zueco2007} but for our purposes it will be appropriate to use the integral form by suitably modifying (\ref{starint1}) to get
\ba \label{star1}
W(\theta,\phi) & = & \left(\frac{2j + 1}{4\pi}\right)^2 \int \rm{Tr} \left(\wh{\Delta}^j(\theta,\phi)\wh{\Delta}^j(\theta^{\pa},\phi^{\pa})\wh{\Delta}^j(\theta^{\pb},\phi^{\pb})\right) \no\\
& & \qquad\times\; \Psi(\theta^{\pa},\phi^{\pa}) \bar{\Psi(\theta^{\pb},\phi^{\pb})} \; \rm{d}\Omega^{\pa}\rm{d}\Omega^{\pb},
\ea
where $\rm{d}\Omega = \sin\theta\rm{d}\theta\rm{d}\phi$ is the invariant measure on $\cal{S}^2$.

All functions defined on a sphere can be expanded in terms of a complete set of spherical harmonics, hence we can write the spin amplitude (\ref{spamp1}) in the form
\be \label{spamp2}
\Psi(\theta,\phi) = \sum^{2j}_{l=0} \sum^l_{m=-l} a_{lm} Y_{l,m}(\theta,\phi),
\ee
where, from the orthogonality property of the spherical harmonics, the coefficients are given by
\be
a_{lm} = \int^{2\pi}_0 \int^\pi_0 \Psi(\theta,\phi) \bar{Y_{l,m}(\theta,\phi)} \sin\theta \; \rm{d}\theta\rm{d}\phi.
\ee
Formula (\ref{spamp2}) suggests that spin amplitudes will transform as tensors under rotations, but we will see that, to the contrary, these amplitudes transform as spinors when $2j$ is odd-integral, in particular when $j = 1/2$.

A double application of (\ref{spamp2}) leads to 
\be \label{spamp3}
|\Psi(\theta,\phi)|^2 = \sum_{l_1,m_1} \sum_{l_2,m_2} a_{l_1m_1} \bar{a}_{l_2m_2} Y_{l_1,m_1}(\theta,\phi) \bar{Y_{l_2,m_2}(\theta,\phi)}.
\ee
By analogy with the case of continuous variables \cite{Bracken2010}, we can call 
\be \label{spamp3a}
|\Psi(\theta,\phi)|^2 = \langle\varphi|\wh{\Delta}^j(\theta,\phi) \; \wh{\varrho} \; \wh{\Delta}^j(\theta,\phi)|\varphi\rangle,
\ee
a generalized spin Husimi function $H(\theta,\phi)$ on the sphere.
In particular, if (up to an unimportant phase) $|\varphi\rangle$ is the Coherent Spin State (CSS) \cite{Klauder1985}
\be \label{spamp3b}
|\varphi\rangle = \sqrt{(2j)!} \sum^j_{m=-j} \frac{[\cos\half\theta]^{j + m} [\sin\half\theta]^{j - m}}{\sqrt{(j + m)!(j - m)!}} e^{-im\phi} |j,m\rangle,
\ee
then (\ref{spamp3a}) becomes the spin Husimi function \cite{Scully1993}.
The product of two spherical harmonics with the same arguments can be written as a linear combination of single spherical harmonics in terms of the $3j$-symbols, hence (\ref{spamp3}) can be re-expressed as
\ba \label{spamp4}
|\Psi(\theta,\phi)|^2 & = & \sum_{l_1,m_1} \sum_{l_2,m_2} \sum_{l,m} \sqrt{\frac{(2l_1+1)(2l_2+1)(2l+1)}{4\pi}} a_{l_1m_1} \bar{a}_{l_2m_2} \no\\
& &  \qquad\times 
   \bpm
      l_1 & l_2 & l \\
      m_1 & m_2 & m
   \epm
   \bpm
      l_1 & l_2 & l \\
      0   & 0   & 0
   \epm
\bar{Y_{l,m}(\theta,\phi)}.
\ea

The Wigner function has a similar expansion, so that for any choice of fixed vector we can write
\be
W(\theta,\phi) = \sum^{2j}_{l=0} \sum^l_{m=-l} b_{lm} Y_{l,m}(\theta,\phi),
\ee
with coefficients
\be
b_{lm} = \int^{2\pi}_0 \int^\pi_0 W(\theta,\phi) \bar{Y_{l,m}(\theta,\phi)} \sin\theta \; \rm{d}\theta\rm{d}\phi.
\ee

The symplectic group of transformations that arises in the case of continuous degrees of freedom, acting on functions of variables $(p,q)$ in a 2f-dimensional phase space, is replaced by the group $SU(2)$ of rotations acting on functions on the phase space of the sphere $\cal{S}^2$.
Rotated spin amplitudes are generated from
\be \label{rot1}
\Psi_R(\theta,\phi) = (\cal{W}(\wh{R}) \star \Psi)(\theta,\phi),
\ee
where the rotation operator is given, for spin-$\half$, by
\be
\wh{R}(\gamma,\alpha,\beta) = e^{-i\gamma\wh{\sigma}_z/2} e^{-i\alpha\wh{\sigma}_y/2} e^{-i\beta\wh{\sigma}_z/2}.
\ee

For the case of spin-$\half$ systems ($j = \half$), setting the constants $\epsilon^{1/2}_l = +1$, the WWS kernel (\ref{kern1}) reduces to
\be \label{kern2}
\wh{\Delta}^{1/2}(\theta,\phi) = (\wh{I} + \sqrt{3}\;\vec{n}\cdot\wh{\sigma})/2,
\ee
where $\wh{I}$ is a $2 \times 2$ unit matrix, $\wh{\sigma}$ are the spin operators
\be \label{pauli}
\wh{\sigma}_x =
   \bpm
      0 & 1 \\
      1 & 0
   \epm
   , \;
\wh{\sigma}_y =
   \bpm
      0 & -i \\
      i & 0
   \epm
   , \;
\wh{\sigma}_z =
   \bpm
      1 & 0 \\
      0 & -1
   \epm
   ,
\ee
and $\vec{n} = (\sin\theta\cos\phi,\sin\theta\sin\phi,\cos\theta)$ is an arbitrary unit vector which is parameterized by the spherical polar coordinates $0\leq\theta\leq\pi,$ $0\leq\phi < 2\pi$.

The spin-$\half$ amplitudes then follow by inserting (\ref{ampop1}) and (\ref{kern2}) into (\ref{spamp1}) to get
\be \label{spamp5}
\Psi(\theta,\phi) = \sqrt{\pi} \sum^1_{l=0} \sum^l_{m=-l} a_{lm} Y_{l,m}(\theta,\phi),
\ee
where the coefficients are given by
\ba \label{coeff1}
a_{00} & = & \psi_{1/2}\bar{\varphi}_{1/2} + \psi_{-1/2}\bar{\varphi}_{-1/2}, \no\\
a_{\text{\tiny{1-1}}} & = & \sqrt{2} \psi_{-1/2}\bar{\varphi}_{1/2}, \no\\
a_{10} & = & \psi_{1/2}\bar{\varphi}_{1/2} - \psi_{-1/2}\bar{\varphi}_{-1/2}, \no\\
a_{11} & = & - \sqrt{2} \psi_{1/2}\bar{\varphi}_{-1/2}.
\ea
Using (\ref{spamp4}), we find for the generalized spin Husimi function
\be \label{spamp8}
|\Psi(\theta,\phi)|^2 = \sqrt{\pi} \sum^1_{l=0} \sum^l_{m=-l} b_{lm} Y_{l,m}(\theta,\phi),
\ee
with the associated coefficients
\ba
b_{00} & = & 1, \no\\
b_{\text{\tiny{1-1}}} & = & \frac{1}{\sqrt{2}} (\psi_{-1/2}\bar{\psi}_{1/2} + \varphi_{-1/2}\bar{\varphi}_{1/2}), \no\\
b_{10} & = & |\psi_{1/2}|^2|\varphi_{1/2}|^2 - |\psi_{-1/2}|^2|\varphi_{-1/2}|^2, \no\\
b_{11} & = & - \frac{1}{\sqrt{2}} (\psi_{1/2}\bar{\psi}_{-1/2} + \varphi_{1/2}\bar{\varphi}_{-1/2}).
\ea
The spin Husimi function follows from (\ref{spamp3b}) by choosing $|\varphi\rangle$ to be the CSS
\be
|\varphi\rangle = \bpm
      \cos\half\theta \\
      e^{-i\phi} \sin\half\theta
   \epm,
\ee
and observing that, in terms of expectation values of the spin operators, the elements of the density matrix are given by
\be
|\psi_{\pm1/2}|^2 = \half(1 \pm \langle\wh{\sigma}_z\rangle), \; \psi_{\mp 1/2} \psi_{\pm 1/2} = \half (\langle\wh{\sigma}_x\rangle \pm i \langle\wh{\sigma}_y\rangle).
\ee
Thus, up on inserting the above into (\ref{spamp3a}) leads to \cite{Scully1993}
\be
|\Psi(\theta,\phi)|^2 = H(\theta,\phi) = \rm{const.} \; (1 + \vec{n}\;\langle\wh{\sigma}\rangle).
\ee

The corresponding Wigner functions are obtained by evaluating (\ref{star1}) with (\ref{kern2}) and (\ref{spamp5}), for any choice of fixed spin vector, to get
\be \label{spamp11}
W(\theta,\phi) = \sqrt{\pi} \sum^1_{l=0} \sum^l_{m=-l} c_{lm} Y_{l,m}(\theta,\phi),
\ee
and its coefficients
\ba \label{coeff2}
c_{00} & = & 1, \no\\
c_{\text{\tiny{1-1}}} & = & \sqrt{2} \psi_{-1/2}\bar{\psi}_{1/2}, \no\\
c_{10} & = & |\psi_{1/2}|^2 - |\psi_{-1/2}|^2, \no\\
c_{11} & = & - \sqrt{2} \psi_{1/2}\bar{\psi}_{-1/2}.
\ea
It is now possible to write down immediately the Wigner functions for any arbitrary spin vector by simply inserting appropriately chosen spin components.

It has been established that spin amplitudes are functions on $\cal{S}^2$, and it only remains to be shown that they do indeed transform as spinors under rotations.
To do this consider a rotation through an angle $\alpha$ about the $y$-axis given by
\be \label{rot2}
\wh{R} = \cos\half\alpha\wh{I} - i\sin\half\alpha\wh{\sigma}_y.
\ee
The image $\cal{W}(\wh{R})$ of this rotation is given by the mapping
\be 
\cal{W}(\cos\half\alpha\wh{I} - i\sin\half\alpha\wh{\sigma}_y)(\theta,\phi) \longrightarrow \cos{\half\alpha} - i \sqrt{3} \sin{\half\alpha} \sin{\theta} \sin{\phi}.
\ee
Inserting this into (\ref{rot1}), and evaluating the star product, leads to an expression of the form
\be
\Psi_R(\theta,\phi) = \sqrt{\pi} \sum^1_{l=0} \sum^l_{m=-l} a_{lm} Y_{l,m}(\theta,\phi),
\ee
for the rotated spin amplitudes, with corresponding coefficients given by
\ba \label{rot3}
a_{00} & = & (\psi_{1/2}\bar{\varphi}_{1/2} + \psi_{-1/2}\bar{\varphi}_{-1/2})\cos\half\alpha - (\psi_{-1/2}\bar{\varphi}_{1/2} - \psi_{1/2}\bar{\varphi}_{-1/2})\sin\half\alpha, \no\\ 
a_{\text{\tiny{1-1}}} & = & \sqrt{2} (\psi_{-1/2}\bar{\varphi}_{1/2}\cos\half\alpha + \psi_{1/2}\bar{\varphi}_{1/2}\sin\half\alpha), \no\\ 
a_{10} & = & (\psi_{1/2}\bar{\varphi}_{1/2} - \psi_{-1/2}\bar{\varphi}_{-1/2})\cos\half\alpha - (\psi_{-1/2}\bar{\varphi}_{1/2} + \psi_{1/2}\bar{\varphi}_{-1/2})\sin\half\alpha, \no\\ 
a_{11} & = & - \sqrt{2} (\psi_{1/2}\bar{\varphi}_{-1/2}\cos\half\alpha - \psi_{-1/2}\bar{\varphi}_{-1/2}\sin\half\alpha).
\ea
Setting $\alpha = 2\pi$ in the above set of coefficients produces a change of sign in $\Psi(\theta,\phi)$.
Therefore, even though the spin amplitudes are functions on $\cal{S}^2$, expressed as linear combinations of spherical harmonics, they nevertheless transform as spinors.

\section{Spin amplitudes on the lattice}
Consider a quantum system characterized by two incompatible physical observables $A$ and $B$ each with an associated finite number of (non-degenerate) eigenvalues $\alpha = \{\alpha_i:i=-j,\cdots ,j\}$ and $\beta = \{\beta_i:i=-j,\cdots ,j\}$ respectively.
From these we construct a lattice consisting of $(2j+1)^2$ points $(\alpha,\beta)$ in the plane.
A generalized WW correspondence rule between observables $\wh{X}_j$ and their corresponding symbols $X(\alpha,\beta)$ on each of the lattice points can be defined by
\bse \label{ww1}
\ba 
\label{ww1a}
\wh{X}_j & = & \frac{1}{2j + 1} \sum_{\alpha\beta} X(\alpha,\beta) \wh{\Delta}(\alpha,\beta), \\
\label{ww1b}
X(\alpha,\beta) & = & \rm{Tr}\left(\wh{X}_j\wh{\Delta}(\alpha,\beta)\right),
\ea
\ese
where $\wh{\Delta}(\alpha,\beta)$ is a  $(2j+1) \times (2j+1)$ lattice kernel matrix defined at the lattice points $(\alpha, \beta)$ and required to satisfy the following properties
\bse \label{sect1ab}
\ba 
\wh{\Delta}(\alpha,\beta) & = & \wh{\Delta}^\dagger(\alpha,\beta) \\
\rm{Tr}(\wh{\Delta}(\alpha,\beta)) & = & 1 \\
\rm{Tr}(\wh{\Delta}(\alpha,\beta)\wh{\Delta}(\alpha^{\pa},\beta^{\pa})) & = & (2j + 1) \;\delta_{\alpha\alpha^{\pa}}\delta_{\beta\beta^{\pa}} \\
\sum_{\alpha\beta} \wh{\Delta}(\alpha,\beta) & = & (2j+1) \;\wh{1}.
\ea
\ese
One possible form for the kernel $\wh{\Delta}(\alpha,\beta)$ that guarantees the above properties are satisfied is \cite{Buot1974, Wootters1987}
\be \label{kern3}
\wh{\Delta}(\alpha,\beta) = \sum_{\alpha^{\pa}} e^{-\frac{2\pi i}{2j+1}\beta\alpha^{\pa}} |\alpha + \alpha^{\pa}/2\rangle\langle\alpha - \alpha^{\pa}/2|, \qquad \alpha^{\pa} = -j, \cdots ,j,
\ee
where the $|\cdot\rangle\langle\cdot|$ are $(2j + 1) \times (2j +1 )$ matrices, whose elements are determined from the properties (\ref{sect1ab}).
Alternatively, the matrix elements of $\wh{\Delta}(\alpha,\beta)$ with respect to the basis $|\beta\rangle$ are found from (\ref{kern3}) and given by the expression
\be \label{kern4}
\langle\beta^{\pa}|\wh{\Delta}(\alpha,\beta)|\beta^{\pb}\rangle = e^{\frac{2\pi i}{2j+1}\alpha(\beta^{\pb} - \beta^{\pa})} \; \sum_{\alpha^{\pa}} e^{-\frac{2\pi i}{2j+1}\alpha^{\pa}[\beta + (\beta^{\pa} + \beta^{\pb})/2]}.
\ee
A double application of (\ref{ww1a}) in conjunction with (\ref{ww1b}), leads to the star product 
\be \label{star2}
(X \star Y)(\alpha,\beta) = \frac{1}{(2j + 1)^2} \sum_{\alpha^{\pa}\alpha^{\pb}\beta^{\pa}\beta^{\pb}} X(\alpha^{\pa},\beta^{\pa}) Y(\alpha^{\pb},\beta^{\pb}) \rm{Tr}\left(\wh{\Delta}(\alpha,\beta) \wh{\Delta}(\alpha^{\pa},\beta^{\pa}) \wh{\Delta}(\alpha^{\pb},\beta^{\pb})\right),
\ee
of two symbols defined on the lattice points.

The image of the amplitude operator $\wh{\Psi}$ of (\ref{ampop1}) under the action of (\ref{ww1b}) gives the amplitude on the lattice as
\be \label{spamp6}
\Psi(\alpha,\beta) = \rm{Tr}\left(\wh{\Psi}\wh{\Delta}(\alpha,\beta)\right),
\ee
with the corresponding Wigner function, obtained by evaluating the star product (\ref{star2}) with (\ref{spamp6}), given by
\be \label{wf2}
W(\alpha,\beta) = \frac{1}{2j + 1} (\Psi \star \bar{\Psi})(\alpha,\beta).
\ee

The lattice associated with spin $j=\half$ consists of an array of four points designated by  $\{(0,0),(0,1),(1,0),(1,1)\}$.
Defined on the lattice is a complex spinor amplitude $\Psi$ and an associated real Wigner function.
The matrix elements for the lattice kernel are in this case obtained directly from (\ref{kern4}) and found to be
\ba \label{kern5}
\wh{\Delta}_{00} & = &  
   \bpm
      1 & \half (1-i) \\
      \half (1+i) & 0
   \epm
   , \; 
\wh{\Delta}_{10} =   
   \bpm
      1 & -\half (1-i) \\
      -\half (1+i) & 0
   \epm
   , \no\\
\wh{\Delta}_{01} & = & 
   \bpm
      0 & \half (1+i) \\
      \half (1-i) & 1
   \epm
   , \;
\wh{\Delta}_{11} = 
   \bpm
      0 & -\half (1+i) \\
      -\half (1-i) & 1
   \epm
   .
\ea
The symplectic Fourier transform of the kernel $\wh{\Delta}(\alpha,\beta)$, is
\be
\wh{D}(\alpha,\beta) = \frac{1}{2j+1} \sum_{\alpha^{\pa}\beta^{\pa}} e^{\frac{2\pi i}{2j+1}\left(\alpha\beta^{\pa} - \alpha^{\pa}\beta\right)} \wh{\Delta}(\alpha^{\pa},\beta^{\pa}),
\ee
and in particular
\be
\wh{D}(\alpha,\beta) = \half\left(\wh{\Delta}(0,0) + (-1)^\alpha\wh{\Delta}(0,1) + (-1)^\beta\wh{\Delta}(1,0) + (-1)^{\alpha + \beta}\wh{\Delta}(1,1)\right),
\ee
which leads to the Pauli operators
\be
\wh{D}(0,0) = \wh{I}, \; \wh{D}(0,1) = \wh{\sigma}_x, \; \wh{D}(1,1) = \wh{\sigma}_y, \; \wh{D}(1,0) = \wh{\sigma}_z.
\ee

Combining (\ref{ampop1}) and (\ref{kern5}) to evaluate (\ref{spamp6}) we have for the spin amplitudes
\ba \label{spamp7}
\Psi(0,0) & = & \bar{\varphi}_{1/2} (\psi_{1/2} + \psi_{-1/2}) + \half (1 + i) (\psi_{1/2}\bar{\varphi}_{-1/2} - \psi_{-1/2}\bar{\varphi}_{1/2}), \no\\
\Psi(0,1) & = & \psi_{-1/2} (\bar{\varphi}_{-1/2} + \bar{\varphi}_{1/2}) + \half (1 - i) (\psi_{1/2}\bar{\varphi}_{-1/2} - \psi_{-1/2}\bar{\varphi}_{1/2}), \no\\
\Psi(1,0) & = & \bar{\varphi}_{1/2} (\psi_{1/2} - \psi_{-1/2}) - \half (1 + i) (\psi_{1/2}\bar{\varphi}_{-1/2} - \psi_{-1/2}\bar{\varphi}_{1/2}), \no\\
\Psi(1,1) & = & \psi_{-1/2} (\bar{\varphi}_{-1/2} - \bar{\varphi}_{1/2}) - \half (1 - i) (\psi_{1/2}\bar{\varphi}_{-1/2} - \psi_{-1/2}\bar{\varphi}_{1/2}).
\ea
These lead immediately to
\ba
|\Psi(0,0)|^2 & = & |(\psi_{1/2} + \psi_{-1/2})\bar{\varphi}_{1/2} + i(\bar{\varphi}_{1/2} + \bar{\varphi}_{-1/2})\psi_{1/2}|^2/2, \no\\
|\Psi(0,1)|^2 & = & |(\varphi_{1/2} + \varphi_{-1/2})\bar{\psi}_{-1/2} + i(\bar{\psi}_{1/2} + \bar{\psi}_{-1/2})\varphi_{1/2}|^2/2, \no\\
|\Psi(1,0)|^2 & = & |(\psi_{1/2} - \psi_{-1/2})\bar{\varphi}_{1/2} + i(\bar{\varphi}_{1/2} - \bar{\varphi}_{-1/2})\psi_{1/2}|^2/2, \no\\
|\Psi(1,1)|^2 & = & |(\varphi_{1/2} - \varphi_{-1/2})\bar{\psi}_{-1/2} + i(\bar{\psi}_{1/2} - \bar{\psi}_{-1/2})\varphi_{1/2}|^2/2,
\ea
which defines a discrete spin Husimi distribution \cite{Marchiolli2009} on the lattice.
Using the amplitudes (\ref{spamp7}) in (\ref{wf2}) it is found that, whatever choice is made for the fixed spin state $|\varphi\rangle$, the Wigner function for any given state $|\psi\rangle$ is
\ba \label{spamp9}
W(0,0) & = & [\bar{\psi}_{1/2} (\psi_{1/2} + \psi_{-1/2}) + \half (1 + i) (\psi_{1/2}\bar{\psi}_{-1/2} - \psi_{-1/2}\bar{\psi}_{1/2})]/2, \no\\
W(0,1) & = & [\psi_{-1/2} (\bar{\psi}_{-1/2} + \bar{\psi}_{1/2}) + \half (1 - i) (\psi_{1/2}\bar{\psi}_{-1/2} - \psi_{-1/2}\bar{\psi}_{1/2})]/2, \no\\
W(1,0) & = & [\bar{\psi}_{1/2} (\psi_{1/2} - \psi_{-1/2}) - \half (1 + i) (\psi_{1/2}\bar{\psi}_{-1/2} - \psi_{-1/2}\bar{\psi}_{1/2})]/2, \no\\
W(1,1) & = & [\psi_{-1/2} (\bar{\psi}_{-1/2} - \bar{\psi}_{1/2}) - \half (1 - i) (\psi_{1/2}\bar{\psi}_{-1/2} - \psi_{-1/2}\bar{\psi}_{1/2})]/2.
\ea
Note that even on a lattice the Wigner function can take negative values and therefore cannot be interpreted as a probability density function.
This can be seen by choosing, for example, $\psi_{1/2} = (1-i)/\sqrt{2}, \psi_{-1/2} = \sqrt{2}(1+i)$.

The images $\cal{W}(\wh{R})$ of a rotation about the $y$-axis, given by (\ref{rot2}), on each of the four lattice points are then given by the relations
\be \label{spamp10}
R(0,0) = R(1,1) = e^{-i\alpha/2}, \;\; R(0,1) = R(1,0) = e^{i\alpha/2}.
\ee
Using (\ref{spamp10}) and (\ref{spamp7}), the rotated spin amplitudes on each of the lattice points are obtained from the formula
\be
\Psi_R(\alpha,\beta) = (R \star \Psi)(\alpha,\beta).
\ee
This is evaluated according to the formula (\ref{star2}) and yields
\ba \label{spamprot1}
\Psi_R(0,0) & = & \bar{\varphi}_{1/2}(\psi_{1/2} + \psi_{-1/2})\cos\half\alpha + \bar{\varphi}_{1/2}(\psi_{1/2} - \psi_{-1/2})\sin\half\alpha + \half(1 + i) \no\\
& & \times ((\psi_{1/2}\bar{\varphi}_{-1/2} - \psi_{-1/2}\bar{\varphi}_{1/2})\cos\half\alpha - (\psi_{1/2}\bar{\varphi}_{1/2} + \psi_{-1/2}\bar{\varphi}_{-1/2})\sin\half\alpha), \no\\
\Psi_R(0,1) & = & \psi_{-1/2}(\bar{\varphi}_{-1/2} + \bar{\varphi}_{1/2})\cos\half\alpha + \psi_{1/2}(\bar{\varphi}_{1/2} + \bar{\varphi}_{-1/2})\sin\half\alpha + \half(1 - i) \no\\
& & \times ((\psi_{1/2}\bar{\varphi}_{-1/2} - \psi_{-1/2}\bar{\varphi}_{1/2})\cos\half\alpha - (\psi_{1/2}\bar{\varphi}_{1/2} + \psi_{-1/2}\bar{\varphi}_{-1/2})\sin\half\alpha), \no\\
\Psi_R(1,0) & = & \bar{\varphi}_{1/2}(\psi_{1/2} - \psi_{-1/2})\cos\half\alpha - \bar{\varphi}_{1/2}(\psi_{1/2} + \psi_{-1/2})\sin\half\alpha - \half(1 + i) \no\\
& & \times ((\psi_{1/2}\bar{\varphi}_{-1/2} - \psi_{-1/2}\bar{\varphi}_{1/2})\cos\half\alpha - (\psi_{1/2}\bar{\varphi}_{1/2} + \psi_{-1/2}\bar{\varphi}_{-1/2})\sin\half\alpha), \no\\
\Psi_R(1,1) & = & \psi_{-1/2}(\bar{\varphi}_{-1/2} - \bar{\varphi}_{1/2})\cos\half\alpha + \psi_{1/2}(\bar{\varphi}_{-1/2} - \bar{\varphi}_{1/2})\sin\half\alpha - \half(1 - i) \no\\
& & \times ((\psi_{1/2}\bar{\varphi}_{-1/2} - \psi_{-1/2}\bar{\varphi}_{1/2})\cos\half\alpha - (\psi_{1/2}\bar{\varphi}_{1/2} + \psi_{-1/2}\bar{\varphi}_{-1/2})\sin\half\alpha). \no\\
& & 
\ea
Setting $\alpha = 2\pi$ in the above set of amplitudes results in a change of sign in $\Psi(\alpha,\beta)$ and thus shows that spin amplitudes defined on the lattice also transform as spinors under rotations.

\section{State Superposition in Phase Space}
The most important new feature that amplitudes bring to the description of quantum systems in phase space is their linear superposability, a property that they inherit from the superposability of quantum state vectors by virtue of the definitions (\ref{spamp1}) and (\ref{spamp6}), which are linear in the relevant state vectors.
Because the definition (\ref{spamp1a}) and (\ref{wf2}) of a Wigner function is not linear in the corresponding state vector, there is no simple way to construct $W$ from $W_1$ and $W_2$ using phase space methods, when $W$, $W_1$ and $W_2$ are Wigner functions corresponding to the pure states $|\psi\rangle$, $|\psi_1\rangle$ and $|\psi_2\rangle$ in the superposition
\be \label{super1}
|\psi\rangle = \alpha |\psi_1\rangle + \beta |\psi_2\rangle.
\ee
However, if the phase space amplitudes $\Psi_1$ and $\Psi_2$ corresponding to $|\psi_1\rangle$ and $|\psi_2\rangle$ are constructed, for one particular choice of reference (window) state vector $|\varphi\rangle$, then we evidently have for the amplitude $\Psi$ corresponding to the superposition (\ref{super1}), simply
\be
\Psi = \alpha \Psi_1 + \beta \Psi_2.
\ee
Then the corresponding Wigner function $W$ can be constructed from the generalized Born interpretation, as
\ba
W = \Psi \star \bar{\Psi} & = & (\alpha \Psi_1 + \beta \Psi_2) \star (\bar{\alpha \Psi_1 + \beta \Psi_2}) \no\\
& = & |\alpha|^2 \Psi_1 \star \bar{\Psi_1} + |\beta|^2 \Psi_2 \star \bar{\Psi_2} + \alpha\bar{\beta} \Psi_1 \star \bar{\Psi_2} + \beta\bar{\alpha} \Psi_2 \star \bar{\Psi_1} \no\\
& = & |\alpha|^2 W_1 + |\beta|^2 W_2 + \alpha\bar{\beta} \Psi_1 \star \bar{\Psi_2} + \beta\bar{\alpha} \Psi_2 \star \bar{\Psi_1}.
\ea
The cross terms in the final expression here emphasize the importance of spin amplitudes, over and above that of Wigner functions, in the phase space description of superposition states.

These considerations apply to amplitudes defined on the sphere or the lattice, just as they do in the continuous case \cite{Bracken2010}.
Similar remarks apply also to the construction of spin Husimi functions: if $H$, $H_1$ and $H_2$ correspond to $|\psi\rangle$, $|\psi_1\rangle$ and $|\psi_2\rangle$ in (\ref{super1}), then there is no straightforward phase space method to construct $H$ from $H_1$ and $H_2$.
However, with the introduction of phase space amplitudes, on the sphere or lattice, we have
\ba
H = |\Psi|^2 & = & |\alpha \Psi_1 + \beta \Psi_2|^2 \no\\
& = & |\alpha|^2 H_1 + |\beta|^2 H_2 + \alpha\bar{\beta} \Psi_1\bar{\Psi_2} + \beta\bar{\alpha} \Psi_2\bar{\Psi_1}.
\ea
Note that $\Psi$ carries the same information about the relative phases of $|\psi_1\rangle$ and $|\psi_2\rangle$ as does the linear combination $|\psi\rangle$.
Interference phenomena will therefore be described transparently in phase space in terms of amplitudes, whereas the description of such phenomena in terms of Wigner or Husimi functions is necessarily less clear.

\section{Example: Spinor Amplitudes in a Magnetic Field}
An important example of a quantum control system involves the manipulation of a spin-$\half$ magnetic dipole with an external magnetic field.
Procedures of this kind are prevalent in nuclear magnetic resonance (NMR) spectroscopy.
In this case non-interacting spin-$\half$ dipoles are immersed in an external homogeneous magnetic field and perturbed from equilibrium by a controlled magnetic field oscillating in the plane perpendicular to the external field.
It is illuminating to investigate the form and behaviour of a spinor amplitude representing an isolated spin-$\half$ dipole under the influence of an NMR control environment.

Consider then such a dipole exposed to an external constant magnetic field $B_z$ acting along the $z$-axis and subjected to a (weaker) controlled magnetic field $B_{xy}$ rotating in the $xy$-axis.
The governing time dependent Hamiltonian at the time $t$ when a phased sinusoidal pulse is applied is given by \cite{Levitt2008}
\be \label{nmr1}
\wh{H} = \half\omega_0 \wh{\sigma}_z + \half\omega_{\rm{nut}} [\cos(\omega_{\rm{ref}} t + \chi_p)\wh{\sigma}_x + \sin(\omega_{\rm{ref}} t + \chi_p)\wh{\sigma}_y],
\ee
where $\omega_0 = -\gamma B_z$ is the Larmor frequency, $\omega_{\rm{nut}} = - \gamma B_{xy}$ is the nutation frequency, $\omega_{\rm{ref}}$ is the spectrometer reference frequency, $\gamma$ the gyromagnetic ratio, $\chi_p$ is the phase angle, and $\wh{\sigma}_x,\wh{\sigma}_y,\wh{\sigma}_z$ are the Pauli matrices as in (\ref{pauli}).

A simplification is obtained by eliminating the time from the Hamiltonian (\ref{nmr1}) by transforming to a frame which is rotating at the Larmor frequency.
Under this transformation, the time independent Hamiltonian at the time of the sinusoidal pulse takes the form
\be \label{nmr2}
\wh{H}_{\rm{rot}} = \half\omega_{\rm{res}}\wh{\sigma}_z + \half\omega_{\rm{nut}}(\wh{\sigma}_x\cos\chi_p + \wh{\sigma}_y\sin\chi_p),
\ee
where $\omega_{\rm{res}} = \omega_0 - \omega_{\rm{ref}}$ is the resonance off-set.
The spinor amplitude derived in this frame will appear to be stationary.

The amplitude operator (\ref{ampop1}), with $j = \half$, in the rotating frame is found from
\be \label{nmr8}
\wh{\Psi}^{\pa} = \wh{U} \wh{\Psi},
\ee
where the evolution operator $\wh{U}$ for this Hamiltonian is
\be
\wh{U} = e^{-i\wh{H}_{\rm{rot}}t},
\ee
and is expressed in matrix form as
\be \label{nmr7}
\wh{U} =  
   \bpm
      \cos\half\alpha - i \omega_{\rm{res}}\sin\half\alpha/\omega_{\rm{eff}} & - i e^{-i\chi_p}\omega_{\rm{nut}}\sin\half\alpha/\omega_{\rm{eff}} \\
      - i e^{i\chi_p}\omega_{\rm{nut}}\sin\half\alpha/\omega_{\rm{eff}} & \cos\half\alpha + i \omega_{\rm{res}}\sin\half\alpha/\omega_{\rm{eff}}
   \epm
   .
\ee
Here the flip angle $\alpha = \omega_{\rm{eff}}t$ and the quantity $\omega_{\rm{eff}} = \sqrt{\omega^2_{\rm{res}} + \omega^2_{\rm{nut}}}$ is the magnitude of the off-resonance rotation frequency that gyrates around the tilted axis.

The spinor amplitude $\Psi(\theta,\phi)$ in the rotating frame can be found by evaluating
\be \label{nmr2a}
(U\star\Psi)(\theta,\phi) \;\rm{or}\; \rm{Tr}(\wh{U}\wh{\Psi}\wh{\Delta}^{1/2}(\theta,\phi)).
\ee
Expressed in terms of spherical harmonics the spinor amplitude has the form (\ref{spamp5}), with coefficients now given from (\ref{nmr2a}) by
\ba \label{nmr3}
a_{00} & = & (\psi_{1/2}\bar{\varphi}_{1/2} + \psi_{-1/2}\bar{\varphi}_{-1/2})\cos\half\alpha - i[(\psi_{1/2}\bar{\varphi}_{1/2} - \psi_{-1/2}\bar{\varphi}_{-1/2})\omega_{\rm{res}} \no\\
& & \qquad\qquad + \; (e^{-i\chi_p}\psi_{-1/2}\bar{\varphi}_{1/2} + e^{i\chi_p}\psi_{1/2}\bar{\varphi}_{-1/2})\omega_{\rm{nut}}]\sin\half\alpha/\omega_{\rm{eff}}, \no\\ 
a_{\text{\tiny{1-1}}} & = & \sqrt{2} (\psi_{-1/2}\bar{\varphi}_{1/2}\cos\half\alpha + i[\psi_{-1/2}\bar{\varphi}_{1/2}\omega_{\rm{res}} \no\\
& & \qquad\qquad - \; e^{i\chi_p} \psi_{1/2}\bar{\varphi}_{1/2}\omega_{\rm{nut}}]\sin\half\alpha/\omega_{\rm{eff}}), \no\\ 
a_{10} & = & (\psi_{1/2}\bar{\varphi}_{1/2} - \psi_{-1/2}\bar{\varphi}_{-1/2})\cos\half\alpha - i[(\psi_{1/2}\bar{\varphi}_{1/2} + \psi_{-1/2}\bar{\varphi}_{-1/2})\omega_{\rm{res}} \no\\
& & \qquad\qquad + \; (e^{-i\chi_p}\psi_{-1/2}\bar{\varphi}_{1/2} - e^{i\chi_p}\psi_{1/2}\bar{\varphi}_{-1/2})\omega_{\rm{nut}}]\sin\half\alpha/\omega_{\rm{eff}}, \no\\ 
a_{11} & = & - \sqrt{2} (\psi_{1/2}\bar{\varphi}_{-1/2}\cos\half\alpha - i[\psi_{1/2}\bar{\varphi}_{-1/2}\omega_{\rm{res}} \no\\
& & \qquad\qquad + \; e^{-i\chi_p} \psi_{-1/2}\bar{\varphi}_{-1/2}\omega_{\rm{nut}}]\sin\half\alpha/\omega_{\rm{eff}}).
\ea

As resonance is approached, that is when $\omega_0 \approx \omega_{\rm{ref}}$, we find $\alpha = \omega_{\rm{nut}} t$ and $\omega_{\rm{eff}} = \omega_{\rm{nut}}$ and the coefficients (\ref{nmr3}) reduce to
\ba \label{nmr4}
a_{00} & = & (\psi_{1/2}\bar{\varphi}_{1/2} + \psi_{-1/2}\bar{\varphi}_{-1/2})\cos\half\alpha \no\\
& & \qquad - i(e^{-i\chi_p}\psi_{-1/2}\bar{\varphi}_{1/2} + e^{i\chi_p}\psi_{1/2}\bar{\varphi}_{-1/2})\sin\half\alpha, \no\\ 
a_{\text{\tiny{1-1}}} & = & \sqrt{2} (\psi_{-1/2}\bar{\varphi}_{1/2}\cos\half\alpha - i e^{i\chi_p} \psi_{1/2}\bar{\varphi}_{1/2}\sin\half\alpha), \no\\ 
a_{10} & = & (\psi_{1/2}\bar{\varphi}_{1/2} - \psi_{-1/2}\bar{\varphi}_{-1/2})\cos\half\alpha \no\\
& & \qquad - i(e^{-i\chi_p}\psi_{-1/2}\bar{\varphi}_{1/2} - e^{i\chi_p}\psi_{1/2}\bar{\varphi}_{-1/2})\sin\half\alpha, \no\\ 
a_{11} & = & - \sqrt{2} (\psi_{1/2}\bar{\varphi}_{-1/2}\cos\half\alpha - i e^{-i\chi_p} \psi_{-1/2}\bar{\varphi}_{-1/2}\sin\half\alpha).
\ea
Thus, at resonance the spinor amplitude is subjected to a phased rotation through an angle $\alpha$, where the amount of rotation is dictated by the duration of the controlled magnetic field.
In particular, when the phase $\chi_p = \pi/2$, then (\ref{nmr4}) are reduced to the coefficients that represent a rotation about the $y$-axis through an angle $\alpha$ (compare with (\ref{rot3})).
By switching off the magnetic field, that is setting $\omega_{\rm{nut}} = 0$ in (\ref{nmr4}), the spinor amplitude returns back to its initial equilibrium position of precessing about the $z$-axis.

In the case of a lattice with an associated spin $j = \half$, the phase space is comprised of an array of four points.
The lattice spinor amplitude for an NMR type control configuration for a single spin-$\half$ system that is described by the rotating frame Hamiltonian (\ref{nmr2}) is obtained by using (\ref{nmr8}) in (\ref{spamp6}) to give
\be \label{nmr9}
\Psi(\alpha,\beta) = \rm{Tr}\left(\wh{U}\wh{\Psi}\wh{\Delta}(\alpha,\beta)\right),
\ee
where $\wh{U}$, as before, is given by (\ref{nmr7}), and the lattice kernel $\wh{\Delta}$ by (\ref{kern5}).
Evaluating (\ref{nmr9}), we find for the lattice spinor amplitude
\ba \label{nmr5}
\Psi(0,0) & = & \bar{\varphi}_{1/2} (\psi_{1/2} + \psi_{-1/2})\cos\half\alpha \no\\
& & + \; (\bar{\varphi}_{-1/2} - i\bar{\varphi}_{1/2})(\psi_{1/2}\omega_{\rm{res}}+ \psi_{-1/2}\omega_{\rm{nut}} e^{-i\chi_p})\sin\half\alpha/\omega_{\rm{eff}} \no\\
& & + \; \half (1 + i) ((\psi_{1/2}\bar{\varphi}_{-1/2} - \psi_{-1/2}\bar{\varphi}_{1/2})\cos\half\alpha \no\\
& & + \; ((\psi_{-1/2}\bar{\varphi}_{1/2} - \psi_{1/2}\bar{\varphi}_{-1/2})\omega_{\rm{res}} \no\\
& & - \; (\psi_{1/2}\bar{\varphi}_{1/2} e^{i\chi_p} + \psi_{-1/2}\bar{\varphi}_{-1/2} e^{-i\chi_p})\omega_{\rm{nut}})\sin\half\alpha/\omega_{\rm{eff}}), \no\\
\Psi(0,1) & = & \bar{\varphi}_{-1/2} (\psi_{1/2} + \psi_{-1/2})\cos\half\alpha \no\\
& & - \; (\bar{\varphi}_{1/2} - \bar{\varphi}_{-1/2})(\psi_{-1/2}\omega_{\rm{res}} - \psi_{1/2}\omega_{\rm{nut}} e^{i\chi_p})\sin\half\alpha/\omega_{\rm{eff}} \no\\
& & + \; \half (1 + i) ((\psi_{-1/2}\bar{\varphi}_{1/2} - \psi_{1/2}\bar{\varphi}_{-1/2})\cos\half\alpha \no\\
& & + \; ((\psi_{-1/2}\bar{\varphi}_{1/2} - \psi_{1/2}\bar{\varphi}_{-1/2})\omega_{\rm{res}} \no\\
& & - \; (\psi_{1/2}\bar{\varphi}_{1/2} e^{i\chi_p} + \psi_{-1/2}\bar{\varphi}_{-1/2} e^{-i\chi_p})\omega_{\rm{nut}})\sin\half\alpha/\omega_{\rm{eff}}), \no\\
\Psi(1,0) & = & \bar{\varphi}_{1/2} (\psi_{1/2} - \psi_{-1/2})\cos\half\alpha \no\\
& & - \; (\bar{\varphi}_{-1/2} + i\bar{\varphi}_{1/2})(\psi_{1/2}\omega_{\rm{res}} + \psi_{-1/2}\omega_{\rm{nut}} e^{-i\chi_p})\sin\half\alpha/\omega_{\rm{eff}} \no\\
& & - \; \half (1 + i) ((\psi_{1/2}\bar{\varphi}_{-1/2} - \psi_{-1/2}\bar{\varphi}_{1/2})\cos\half\alpha \no\\
& & - \; ((\psi_{1/2}\bar{\varphi}_{-1/2} - \psi_{-1/2}\bar{\varphi}_{1/2})\omega_{\rm{res}} \no\\
& & + \; (\psi_{1/2}\bar{\varphi}_{1/2} e^{i\chi_p} + \psi_{-1/2}\bar{\varphi}_{-1/2} e^{-i\chi_p})\omega_{\rm{nut}})\sin\half\alpha/\omega_{\rm{eff}}), \no\\
\Psi(1,1) & = & \bar{\varphi}_{-1/2}(\psi_{-1/2} - \psi_{1/2})\cos\half\alpha \no\\
& & + \; (\bar{\varphi}_{1/2} + \bar{\varphi}_{-1/2})(\psi_{-1/2}\omega_{\rm{res}} - \psi_{1/2}\omega_{\rm{nut}} e^{i\chi_p})\sin\half\alpha/\omega_{\rm{eff}} \no\\
& & + \; \half(1 + i) ((\psi_{1/2}\bar{\varphi}_{-1/2} - \psi_{-1/2}\bar{\varphi}_{1/2})\cos\half\alpha \no\\
& & + \; ((\psi_{1/2}\bar{\varphi}_{-1/2} - \psi_{-1/2}\bar{\varphi}_{1/2})\omega_{\rm{res}} \no\\
& & + \; (\psi_{1/2}\bar{\varphi}_{1/2} e^{i\chi_p} + \psi_{-1/2}\bar{\varphi}_{-1/2} e^{-i\chi_p})\omega_{\rm{nut}})\sin\half\alpha/\omega_{\rm{eff}}),
\ea
where the flip angle $\alpha = \omega_{\rm{eff}}t$ and $\omega_{\rm{eff}} = \sqrt{\omega^2_{\rm{res}} + \omega^2_{\rm{nut}}}$.
When $\omega_0 \approx \omega_{\rm{ref}}$, near resonance, we find that (\ref{nmr5}), after a slight rearrangement, reduce to
\ba \label{nmr6}
\Psi(0,0) & = & \bar{\varphi}_{1/2} (\psi_{1/2} + \psi_{-1/2})\cos\half\alpha + \psi_{-1/2} e^{-i\chi_p}(\bar{\varphi}_{-1/2} - i\bar{\varphi}_{1/2})\sin\half\alpha \no\\
& & + \; \half (1 + i) ((\psi_{1/2}\bar{\varphi}_{-1/2} - \psi_{-1/2}\bar{\varphi}_{1/2})\cos\half\alpha \no\\
& & - \; (\psi_{1/2}\bar{\varphi}_{1/2} e^{i\chi_p} + \psi_{-1/2}\bar{\varphi}_{-1/2} e^{-i\chi_p})\sin\half\alpha), \no\\
\Psi(0,1) & = & \psi_{-1/2} (\bar{\varphi}_{-1/2} + \bar{\varphi}_{1/2})\cos\half\alpha + \bar{\varphi}_{-1/2}(\psi_{-1/2} e^{-i\chi_p} + i\psi_{1/2} e^{i\chi_p})\sin\half\alpha \no\\
& & + \; \half (1 - i) ((\psi_{1/2}\bar{\varphi}_{-1/2} - \psi_{-1/2}\bar{\varphi}_{1/2})\cos\half\alpha \no\\
& & + \; (\psi_{1/2}\bar{\varphi}_{1/2} e^{i\chi_p} + \psi_{-1/2}\bar{\varphi}_{-1/2} e^{-i\chi_p})\sin\half\alpha), \no\\
\Psi(1,0) & = & \bar{\varphi}_{1/2} (\psi_{1/2} - \psi_{-1/2})\cos\half\alpha - \psi_{-1/2} e^{-i\chi_p}(\bar{\varphi}_{-1/2} + i\bar{\varphi}_{1/2})\sin\half\alpha \no\\
& & - \; \half (1 + i) ((\psi_{1/2}\bar{\varphi}_{-1/2} - \psi_{-1/2}\bar{\varphi}_{1/2})\cos\half\alpha \no\\
& & - \; (\psi_{1/2}\bar{\varphi}_{1/2} e^{i\chi_p} + \psi_{-1/2}\bar{\varphi}_{-1/2} e^{-i\chi_p})\sin\half\alpha), \no\\
\Psi(1,1) & = & \psi_{-1/2}(\bar{\varphi}_{-1/2} - \bar{\varphi}_{1/2})\cos\half\alpha + \bar{\varphi}_{-1/2}(\psi_{-1/2} e^{-i\chi_p} - i\psi_{1/2} e^{i\chi_p})\sin\half\alpha \no\\
& & - \; \half(1 - i) ((\psi_{1/2}\bar{\varphi}_{-1/2} - \psi_{-1/2}\bar{\varphi}_{1/2})\cos\half\alpha \no\\
& & + \; (\psi_{1/2}\bar{\varphi}_{1/2} e^{i\chi_p} + \psi_{-1/2}\bar{\varphi}_{-1/2} e^{-i\chi_p})\sin\half\alpha).
\ea
At resonance the lattice spinor amplitude experiences a phased rotation through an angle $\alpha$, which is determined by the length of time that the magnetic field is applied.
Choosing a phase angle of $\chi_p = \pi/2$ in (\ref{nmr6}) results in a spinor amplitude representing a rotation about the $y$-axis through an angle $\alpha$ as in (\ref{spamprot1}).
By turning off the magnetic field, that is by setting $\omega_{\rm{nut}} = 0$, the lattice spinor amplitude returns to its equilibrium state.

\section{Concluding remarks}
We have extended the concept of phase space amplitudes to finite spin systems and introduced the notion of a spin amplitude to represent pure spin states in phase space.
Spin amplitudes and their associated Wigner functions have been expressed as linear combinations of spherical harmonics on the sphere.
In addition, a generalized Weyl correspondence has been adapted to define spin amplitudes and Wigner functions on a lattice phase space.
We have concentrated on amplitudes for a single fixed spin, in particular a spin-$\half$ system.
Our approach can be extended in the obvious way to define amplitudes for multipartite spin systems on Cartesian products of one-particle phase spaces, either spheres or lattices.

As an example the general theory has been applied in detail to the case of spin-$\half$, revealing some important features of the approach.
The liberty to choose any normalized fixed spin reference state $|\varphi\rangle$ introduces a degree of arbitrariness into the definition of the spin amplitudes, analogous to the freedom to choose a ``window state" in the continuous case \cite{Bracken2010}.
It is an important problem to optimize this choice for a given quantum spin system.
We emphasize however, as can be seen from (\ref{born1}) and (\ref{wf2}), that the associated Wigner functions are independent of this choice, just as in the continuous case.

The spin amplitudes are completely described in terms of a combination of system and fixed state spinor components and these amplitudes transform as spinors under rotations on both the sphere and lattice thus further supporting their fundamental status. 
Spin amplitudes in phase space can be superposed like state vectors, unlike Wigner functions or Husimi functions.
They enable an expanded phase space description of quantum spin systems, as in our example of a magnetic dipole in a time-dependent magnetic field.

It is our view that the representation of spin states in phase space holds the promise of new physical insights and a novel perspective in areas like quantum computing, quantum control systems, and quantum information. 
We hope to return to some of these aspects in the future.

\end{document}